\documentclass[a4paper,11pt, notitlepage]{article}
\usepackage{geometry}
\usepackage{graphicx}
\usepackage{setspace}
\usepackage{amsfonts}
\usepackage[bbgreekl]{mathbbol}
\DeclareSymbolFontAlphabet{\mathbb}{AMSb}
\DeclareSymbolFontAlphabet{\mathbbl}{bbold}

\usepackage[english]{babel}
\usepackage{caption}
\usepackage{subcaption} 
\usepackage{mathtools}
\usepackage{mathbbol}
\usepackage{enumerate}
\usepackage{xcolor}

\usepackage{array}

\usepackage{amsmath}
\usepackage{graphicx}
\usepackage{graphics}
\usepackage{amssymb}
\usepackage{amsthm}
\usepackage{thmtools, thm-restate}

\newtheorem{theorem}{Theorem}[section]
\newtheorem{definition}[theorem]{Definition}

\newcommand{\ceil}[1]{\left\lceil #1 \right\rceil}

\newcommand{\T}{{t^*}}

\newcommand{\G}{\mathcal{G}}

\newcommand{\RT}{\mathcal{T}}

\newcommand{\N}{\mathbb{N}}

\title{Brief Announcement: \\ Broadcasting Time in Dynamic Rooted Trees is Linear}

\author{Antoine El-Hayek\\Faculty of Computer Science\\UniVie Doctoral School Computer Science DoCS\\University of Vienna, Austria
\and
Monika Henzinger\\Faculty of Computer Science\\University of Vienna, Austria\\
\and
Stefan Schmid\\Faculty of Computer Science\\University of Vienna, Austria \\ TU Berlin, Germany
}
\date{}

\begin{document}
\maketitle
\begin{abstract}
We study the broadcast problem on dynamic networks with $n$ processes. The processes communicate in synchronous rounds along an arbitrary rooted tree. The sequence of trees is given by an adversary whose
 goal is to maximize the number of rounds until at least one process reaches all other processes.
 Previous research has shown a $\lceil{\frac{3n-1}{2}}\rceil-2$ lower bound and an $O(n\log\log n)$  upper bound.
We show the first linear upper bound for this problem, namely $\lceil{(1 + \sqrt 2) n-1}\rceil \approx 2.4n$. Our result follows from a detailed analysis of the evolution of the adjacency matrix of the network over time.
\end{abstract}

\section{Introduction}

Broadcast is one of the most fundamental tasks in distributed systems. The problem comes in many flavors and features intriguing connections to other classic problems such as consensus.  
In general, for many distributed tasks, it is essential to understand the patterns of the spread of information in a network over time.

This paper considers the broadcast problem on \emph{dynamic} networks: communication networks that evolve over time, e.g., due to failures or mobility. In particular, we consider a network of $n$ processes which communicate in synchronous rounds along a sequence of arbitrary rooted trees. The trees are chosen by an adversary, which aims to maximize the broadcast time: the number of rounds it takes until there exists a process that everyone has heard of. 

The time complexity of this basic broadcast problem is non-trivial to analyze and has been an open question for several years. Results from Charron-Bost and Schiper in 2009~\cite{charron2009heard} and Charron-Bost, F{\"u}gger, and Nowak in 2015~\cite{charron2015approximate} imply an $n \log n$ upper bound. In 2019, Zeiner, Schwarz, and Schmid~\cite{schwarz2017linear} gave a linear upper bound when the adversary is restricted to trees with either a constant number of leaves or a constant number of inner nodes. They also gave a $\ceil{\frac{3n-1}{2}}-2$ lower bound. In 2020, F{\"u}gger, Nowak, and
Winkler~\cite{fugger2020radius} improved the general upper bound to $2n\log\log n + O(n)$. So far, it has been an open conjecture~\cite{schwarz2017linear} whether the broadcast time is linear for arbitrary sequences of rooted trees.
 

\section{Model}

We consider the following broadcasting problem. Let $n$ be the number of processes, i.e., the number of nodes in the network. 

\begin{definition}
If $A=([n], E_1)$ and $B=([n], E_2)$ are two directed networks on $n$ nodes, then the \emph{product graph} $A\circ B$ is the network on $n$ nodes, with edge set $E$, where $(x,y) \in E$ if and only if there exist a node $z$ such that $(x,z) \in E_1$ and $(z,y) \in E_2$.
\end{definition}
The \emph{broadcasting problem} considered in this paper is defined as follows:
At each round $t=1,2, \hdots$, an adversary chooses a directed network $G_t$ from an unchanging set of networks $\G$. 
Let $G(t)$ be the product graph $G(t)=G_1\circ \hdots \circ G_t$. 
We define broadcast time $\T$ as the smallest round $t$ where there exists a node in $G(t)$ with an out-edge to every other node. Note that this corresponds to 
 a node that has broadcast its piece of information to everyone.

\begin{definition}
The broadcast time $\T$ of a sequence of graphs $G_1, G_2, \hdots$, is defined as follows:

$$
\T(G_1, G_2, \hdots)=\min\{t \in \N: \exists x \in [n], \forall y \in [n], (x,y)\in G_1\circ\hdots \circ G_t\}
$$

\end{definition}

The goal of the adversary is to make $\T$ as large as possible. 

\begin{definition}
The broadcast time $\T$ of an adversary $\G$, is defined as follows:

$$
\T(\G)=\max\{\T(G_1, G_2, \hdots): \forall i \in \N, G_i \in \G\}
$$
\end{definition}

We restrict the adversary to only select rooted trees over $n$ vertices with self loops and call the resulting problem \emph{broadcasting problem with dynamic rooted trees}. Let us denote this set of trees by $\G=\RT_n$. This means that every directed network in $\G$ is a rooted tree, to which one self-loop is added to each node. The self-loops ensure that no process forgets any piece of information in any round, and the rooted tree ensures broadcast in a finite number of rounds.\footnote{If the adversary can choose a non-rooted graph, it could repeat this graph indefinitely, preventing broadcast.} Our goal is to determine $\T(\RT_n)$. 

Note that even in the simple case where the adversary gives the same directed tree in each round, the broadcast time can be as large as $n-1$, namely if the tree is simply a path. Conversely, in each round, it is easy to see that at least one new edge appears in the product graph, and thus broadcast time cannot be larger than $n^2$. This raises the question how large the broadcast time can be made if in each round a different directed tree can be used.

\section{Contribution}

We settle the open problem about time complexity of broadcast in dynamic
networks, by showing that it is \emph{linear}. Hence, Zeiner et al.'s~\cite{schwarz2017linear}
conjecture is true.
In particular, we present an upper bound of $\ceil { (1+\sqrt{2})n-1}$. Combined with the lower bound given by \cite{schwarz2017linear}, this yields the following theorem:

\begin{theorem}
The broadcasting time in dynamic rooted trees is linear:

$$\ceil{\frac{3n-1}{2}}-2 \leq \T(\RT_n) \leq \ceil { (1+\sqrt{2})n-1}$$
\end{theorem}

Our analysis is enabled by a novel perspective on the problem: adjacency matrices with boolean entries. We analyse how these adjacency matrices evolve over rounds.
We believe this perspective can be useful also for other problems, such as consensus, gossiping, or the broadcasting problem in other settings.

\begin{figure}

\begin{center}
\begin{tabular}{|c c c c|} 
 \hline
 Trivial & \cite{schwarz2017linear} & \cite{fugger2020radius} & New \\ [0.5ex] 
 \hline
 $n^2$ & $n\log n$ & $O(n\log\log n)$ & $(1+\sqrt 2) n$ \\ 
 
  & $k$ leaves: $O(kn)$ &  &  \\
  
   & $k$ inner nodes: $O(kn)$ &  &  \\
   \hline
   
\end{tabular}
\end{center}

    \caption{Previously known and new upper-bounds. "$k$ leaves" refers to the case where the adversary is restricted to trees that have $k$ leaves in each round, whereas "$k$ inner nodes" refers to the case where the adversary is restricted to trees that have $k$ inner nodes.}
    \label{fig:my_label}
\end{figure}

\section{Related Work}\label{sec:relwork}

Broadcasting, gossiping, and other information dissemination problems have been studied by the distributed computing community for decades already~\cite{hedetniemi1988survey}. Most classic literature on network broadcast considers a static setting, e.g., where in each round each node can send information to one neighbor~\cite{hromkovivc1996dissemination}. This model has also been explored in the context of gossiping, e.g., by Fraigniaud and Lazard~\cite{fraigniaud1994methods}.  Kuhn, Lynch and Oshman~\cite{kuhn2010distributed} explore the all-to-all data dissemination problem (gossiping) in an undirected dynamic network, where processes do not know beforehand the total number of processes and must decide on that number. Broadcast has also been studied in dynamic communication networks which evolve randomly, e.g., by Clementi et al.~\cite{clementi2013rumor}, and in the radio network model~\cite{ellen2021constant}, just to give a few examples. 

A closely related yet different problem to broadcasting is the consensus problem. This problem builds up on the heard-of model first introduced by Charron-Bost and Schiper~\cite{charron2009heard}. The authors prove results for the solvability of consensus over a wide range of adversaries. Among other results, they give a $\log n$ upper bound for nonsplit graphs, which are graphs for which every pair of nodes has a common in-neighbor. This would result in an $n\log n$ upper bound for rooted trees when combining it with the result of Charron-Bost, F{\"u}gger and Nowak~\cite{charron2015approximate}. Coulouma, Godard and Peters~\cite{coulouma2015characterization} characterize on which dynamic graphs consensus is solvable, based on broadcastability. Another similar problem is agreement, considered by Santoro and Widmayer~\cite{santoro1989time}, where only a $k$-majority should agree on a value, as opposed to everyone for consensus.

We have studied the broadcasting problem on directed dynamic networks, with an adversary that can choose the communication network at each round among rooted trees. Zeiner, Schwarz, and Schmid~\cite{schwarz2017linear} give a $n\log n$ upper bound to our exact problem by using graph-theoretic reasoning. They also give a $\ceil{\frac{3n-1}{2}}-2$ lower bound by providing an explicit example. They further show that under an adversary that can only choose rooted trees with a fixed number of leaves or internal nodes, broadcast time is linear.

There has also been interest in a problem variant which only differs in the pool of networks the adversary can choose a network from for each communication round. F{\"u}gger, Nowak, and Winkler~\cite{fugger2020radius} give an $O( \log\log n)$ upper bound if the adversary can only choose nonsplit graphs. Combined with the result of Charron-Bost, F{\"u}gger, and Nowak~\cite{charron2015approximate} that states that one can simulate $n-1$ rounds of rooted trees with a round of a nonsplit graph, this gives the previous $O(n\log\log n)$ upper bound for our problem. Dobrev and Vrto~\cite{dobrev2002optimal, dobrev1999optimal} give specific results when the adversary is restricted to hypercubic and tori graphs with some missing edges.

\section{Future Work}\label{sec:conclusion}

Our work opens several avenues for future research. A first open problem is to close the gap between the upper and lower bound. A next interesting question is whether the matrix perspective can yield results for similar problems as well, such as consensus, gossiping or broadcasting. We believe that our approach can be extended to analyze different adversaries, for example,  the setting where the adversary is bound to nonsplit graphs (which are graphs where every pair of nodes has a common in-neighbor), as well as non-adversarial environments.

\section*{Acknowledgments}

This project has received funding from the European Research Council (ERC) under the European Union’s Horizon 2020 research and innovation programme (grant agreement No. 101019564).
This work was further supported by the Austrian Science Fund (FWF) and netIDEE SCIENCE project P 33775-N, and by the Federal Ministry of Education and Research (BMBF, Germany), 6G-RIC under Grant 16KISK020K.
%
We would like to thank Kyrill Winkler for his inputs and feedback on this paper.

\hfill
\includegraphics[scale=0.4]{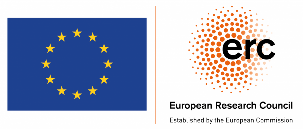}

\bibliographystyle{plain}
\bibliography{broadcastingbib}

\end{document}